\documentstyle[multicol,prb,aps]{revtex}
\begin{document}

\draft
%\twocolumn[

\title{Transport through an Interacting Quantum Dot 
       Coupled to Two Superconducting Leads}

\author{Kicheon Kang\cite{kang}}

\address{ Department of Physics,
          Korea University, Seoul 136-701, Korea  }

\date{\today}

\maketitle

\begin{abstract}
We derive a formula for the current through an interacting quantum dot 
coupled to two supercouducting leads, using the non-equilibrium Green's
function formalism. It is shown that the formula takes an especially simple
form, when the Andreev reflections in the junctions are negligible
because of large Coulomb repulsion in the dot. Our formalism
provides a new framework to investigate many-body effects of mesoscopic systems
in the presence of the superconducting leads. For an example, we describe the 
resonant tunneling of quasiparticles in the presence of strong Coulomb 
repulsion in the dot. Further, we discuss the boson assisted transport.
Multiple sharp peaks are found in the current-voltage curve, which originate 
from the singularity of BCS density of states and the 
boson assisted tunneling. 
\end{abstract}
%]
%\narrowtext
%
\begin{multicols}{2}
Recent advancement of modern nanofabrication technique has stimulated
a lot of interest in the study of electron transport
in quantum dot devices \cite{kastner,ralph94}. 
In addition to the Coulomb blockade
and the resonant tunneling, a novel non-equilibrium Kondo effect
is expected to occur in the low temperature regime for
an ultra small quantum dot or artificial atom which is coupled
to two metallic leads 
\cite{hershfeld,meir91,meir,yeyati93,ng96,stafford96,konig96,kang97}.
More recently, transport through a nanoparticle is studied 
experimentally, where the particle is connected to superconducting 
(Al) leads rather than normal metallic leads \cite{ralph}.
The level spacing of a ``superconducting" nanoparticle
can be either larger or smaller than its bulk superconducting
energy gap, depending on the size of the particle and other
parameters such as temperature or external magnetic field.
When the level spacing of the nanoparticle is sufficiently
large, the superconducting order in the dot disappears and the particle
can be regarded as an ordinary atom \cite{anderson59}.

In this paper, we present a formulation of calculating the current 
through an interacting quantum dot with single level
coupled to two superconducting leads (Fig. 1).
The superconducting leads are characterized by the energy gaps $\Delta_L$,
$\Delta_R$, and the chemical potentials $\mu_L$, $\mu_R$, where 
$\mu_L-\mu_R=eV$, $V$ being the bias voltage across the leads.
The subscript $L(R)$ represents the left (right) lead.
Using the non-equilibrium Green's function technique, we derive an
expression of the steady state current due to quasiparticle processes.
It is shown that the current can be expressed
in terms of the single-particle spectral density of states of the
dot electron, if the Andreev reflection (AR) processes are negligible.
In fact, the AR contributions to the current can be neglected when the Coulomb 
interaction energy $(U)$ in the dot exceeds both
the superconducting gaps $\Delta_L,\Delta_R$
and the coupling between the dot and the leads
$\Gamma_L,\Gamma_R$. It is because, on this condition,
the $2e$-charge fluctuation of the dot is energetically unfavourable. 
Note that this will be a typical condition of double barrier tunneling 
structures with sufficiently small quantum dot (see e.g.~\cite{ralph}). 
In this case the formulation becomes quite simple, because there is no 
AC component of the current.  

To describe the system, we start with the following model Hamiltonian
\begin{eqnarray}
 {\cal H} &=& {\cal H}_L + {\cal H}_R + {\cal H}_D\{d_{\sigma}^{\dagger};
            d_{\sigma}\}  \nonumber \\
 &+& \sum_{k\sigma \in L,R} V_k \left( c_{k\sigma}^{\dagger}d_{\sigma} +
   d_{\sigma}^{\dagger}c_{k\sigma} \right) , \label{eq:hamil}
\end{eqnarray}
where ${\cal H}_L$, ${\cal H}_R$ and ${\cal H}_D$ represent left, right
BCS superconducting leads and the quatum dot, respectively. The last term 
in the Eq.~(\ref{eq:hamil}) corresponds to the coupling between the dot
and each of the leads. $d_{\sigma}^{\dagger}(d_{\sigma})$ creates (destroys)
an electron with spin $\sigma$ in the dot, and $c_{k\sigma}^{\dagger}
(c_{k\sigma})$ creates (destroys) an electron with the momentum $k$ and 
spin $\sigma$ in either the left (L)
or the right (R) lead. In general, ${\cal H}_D$ includes intra-dot
electron-electron interactions, and coupling to an
environment like a bosonic bath through the electron-boson
interactions\cite{konig96}.

As noted above, when $U\gg \Delta_{L,R},\Gamma_{L,R}$ the Andreev reflection
processes are prohibited, and DC current flows by the resonant
tunneling of quasiparticles.
The current flowing out of the lead $\alpha$ ($\alpha=L,R$) 
is written by
\begin{equation}
 I_{\alpha} = \frac{e}{\hbar}\sum_{k\sigma\in\alpha} V_k \int_{-\infty}^{\infty}
     \frac{ d\omega }{ 2\pi } \left( G_{d,k\sigma}^<(\omega) - 
     G_{k\sigma,d}^<(\omega)  \right) , \label{eq:curra0}
\end{equation}
where $G_{d,k\sigma}^<(\omega)$ is the Fourier transform of the Keldysh
Green's function $G_{d,k\sigma}^<(t) \equiv i\left\langle c_{k\sigma}^{\dagger}
(0) d_{\sigma}(t) \right\rangle$. 
By using the Dyson's equation in the Keldysh's formalism,
the current due to quasiparticles is given by    
\begin{eqnarray}
 I_{\alpha} &=& \frac{e}{\hbar}\sum_{k\sigma\in\alpha} 
      V_k^2 \int_{-\infty}^{\infty}
      \frac{ d\omega }{ 2\pi }    \left[
      \left( g_{k\sigma}^<(\omega)-g_{k\sigma}^>(\omega) \right)
      G_{\sigma}^<(\omega)  \right. \nonumber \\  
     & & + \left.    g_{k\sigma}^<(\omega) \left(
      G_{\sigma}^r(\omega) - G_{\sigma}^a(\omega)      \right)
      \right]   ,  \label{eq:curra}
\end{eqnarray}
where $G_{\sigma}^<(t) \equiv i\left\langle d_{\sigma}^{\dagger}(0)
d_{\sigma}(t) \right\rangle$, and $G_{\sigma}^r$ and $G_{\sigma}^a$ are
the corresponding retarded and advanced Green's functions, respectively. 
The Green's functions denoted by $g$ are the unperturbed Green's functions
of the superconducting lead.
In deriving the Eq.(\ref{eq:curra}) from the Eq.(\ref{eq:curra0}), we
have neglected terms containing the off-diagonal Green's functions 
of dot electron in the Nambu representation because those terms 
should involve Andreev reflection processes, which is negligible in our case.
It should be noted that the ``full" Green's functions of the dot represented
by $G_{\sigma}$ may include the Andreev reflections, but
contribution of those can be neglected when the charging energy 
in the dot is sufficiently large. Therefore the Eq.~(\ref{eq:curra}) 
can be adopted as the formula for the current
in the system.

Using the relations
\begin{mathletters}
\begin{eqnarray}
 g_{k\sigma}^<(\omega) &=& 2\pi i \left[ u_k^2 f_{\alpha}(E_k) \delta(\omega-E_k)
       \right. \nonumber \\
    & & \left. + v_k^2\left( 1-f_{\alpha}(E_k) \right) \delta(\omega+E_k) \right] , \\
  g_{k\sigma}^>(\omega) &=& -2\pi i \left[ u_k^2 \left(1-f_{\alpha}(E_k)\right)
     \delta(\omega-E_k) \right]  \nonumber \\
    & & \left. + v_k^2 f_{\alpha}(E_k) \delta(\omega+E_k) \right] ,
\end{eqnarray}
\end{mathletters}
where $u_k^2=\frac{1}{2}\left( 1+\frac{ \epsilon_k-\mu_{\alpha} }{ E_k-\mu_{\alpha} } 
\right)$, $v_k^2=\frac{1}{2}\left( 1-\frac{ \epsilon_k-\mu_{\alpha} }{ E_k-\mu_{\alpha} } 
\right)$, 
$\epsilon_k$ represents the (normal) single particle energy of the
lead, $E_k=\mu_{\alpha} \pm\sqrt{ (\epsilon_k-\mu_{\alpha})^2 + 
\Delta_{\alpha}^2 }$, and 
$f_{\alpha}$ is the Fermi distribution function of the lead $\alpha$,
%
%Using the expression for the unperturbed superconducting Green's functions
%$g_{k\sigma}^<(\omega), g_{k\sigma}^>(\omega)$,
%
one find
\begin{eqnarray}
 I_{\alpha} &=& \frac{2ie}{h}\sum_{\sigma}\int d\omega\, 
     \Gamma_{\alpha}^S(\omega) \nonumber \\
     & & \times \left[
       f_{\alpha}(\omega) \left( G_{\sigma}^r(\omega)-G_{\sigma}^a(\omega) 
       \right)
       + G_{\sigma}^<(\omega)      \right] .
	 \label{eq:curra2}
\end{eqnarray}
Here $\Gamma_{\alpha}^S(\omega) = \Gamma_{\alpha} 
\zeta_{\alpha}(\omega-\mu_{\alpha})$, 
with $\Gamma_{\alpha} = \pi\sum_{k\in\alpha}
V_k^2\,\delta(\mu_{\alpha}-\epsilon_k)$ 
being the constant characterizing coupling between
the dot and the lead $\alpha$ which is assumed
to be independent of the energy, and $\zeta_{\alpha}(\omega)$ being the 
dimensionless BCS density of states of the lead $\alpha$:
$\zeta_{\alpha}(\omega) = |\omega|/\sqrt{\omega^2-
\Delta_{\alpha}^2 }$ if $|\omega| > \Delta_{\alpha}$, and otherwise 0.
It is straightforward to generallize the Eq.~(\ref{eq:curra2}) to
the system with a quantum dot containing multi-levels.
In that case, $\Gamma_{\alpha}^S$ and the Green's functions are replaced by
$M\times M$ matrices, $M$ being the number of the levels,
and then $I_{\alpha}$ is given by the trace of the 
$M\times M$ matrix in the expression~(\ref{eq:curra2}).
In the steady state, the current is conserved,
that is $I_L+I_R=0$, and 
thus one can symmetrize the formula for the current as
\begin{eqnarray}
 I &=& I_L = -I_R = \frac{1}{2}\left( I_L-I_R \right) \nonumber \\ 
   &=& \frac{ie}{h}\sum_{\sigma}\int d\omega\,  \left\{ \left( \Gamma_L^S(\omega)f_L(\omega)
       -\Gamma_R(\omega)f_R(\omega) \right) \right. \label{eq:curr} \\
   & & \left. \times    \left( G_{\sigma}^r(\omega)-G_{\sigma}^a(\omega) \right)
        + \left( \Gamma_L^S(\omega)-\Gamma_R^S(\omega) \right) G_{\sigma}^<(\omega)
	\right\}  ,  \nonumber
\end{eqnarray}
with the condition of the current conservation $(I_L=-I_R)$, 
\begin{eqnarray}
   &  \int d\omega\, \left( \Gamma_L^S(\omega)+\Gamma_R^S(\omega) \right) G_{\sigma}^<(\omega)
     & \nonumber \\
   & =  -\int d\omega\, \left( \Gamma_L^S(\omega)f_L(\omega)+\Gamma_R^S(\omega)f_R(\omega) \right)
     & \label{eq:steady} \\
   &  \times \left( G_{\sigma}^r(\omega) - G_{\sigma}^a(\omega) \right). 
     & \nonumber
\end{eqnarray}
The current formula (\ref{eq:curr}) is reduced to the simple form
\begin{equation}
 I = \frac{2e}{\hbar} \sum_{\sigma}\int d\omega\, \tilde{\Gamma}^S(\omega)
     \left\{ f_L(\omega)-f_R(\omega) \right\} \rho_{\sigma}(\omega) ,
	       \label{eq:curr2}
\end{equation}
where $\tilde{\Gamma}^S(\omega) = \frac{ \Gamma_L^S(\omega)\Gamma_R^S(\omega) }{
\Gamma_L^S(\omega)+\Gamma_R^S(\omega) }$ and $\rho_{\sigma}(\omega)=-\frac{1}{\pi}
\mbox{Im}\,G_{\sigma}^r(\omega)$, by using the approximation
\begin{eqnarray}
 & & \int d\omega\, \left( \Gamma_L^S(\omega)-\Gamma_R^S(\omega) \right) G_{\sigma}^<(\omega) 
      \nonumber \\
 & & = -\int d\omega\, \frac{ \Gamma_L^S(\omega)-\Gamma_R^S(\omega) }{
     \Gamma_L^S(\omega)+\Gamma_R^S(\omega) }
       \\
 & & \times \left( \Gamma_L^S(\omega)f_L(\omega)+\Gamma_R^S(\omega)f_R(\omega) \right)
     \left( G_{\sigma}^r(\omega) - G_{\sigma}^a(\omega) \right)  \nonumber .
\end{eqnarray}
This approximation is made by multiplying the factor 
$\frac{ \Gamma_L^S(\omega)-\Gamma_R^S(\omega) }{ \Gamma_L^S(\omega)+\Gamma_R^S(\omega) }$ on 
the integrands of both sides of the Eq.~(\ref{eq:steady}), and seems to be
reasonable since $\frac{ \Gamma_L^S(\omega)-\Gamma_R^S(\omega) }{ 
\Gamma_L^S(\omega)+\Gamma_R^S(\omega) }$ is a smooth function of $\omega$ except for the
{\em points} $\omega=\mu_{\alpha}\pm\Delta_{\alpha}$ and its absolute value
is not larger than one. 
Further, this approximation becomes exact in the small coupling limit,
$\Gamma\ll k_BT$, because both $G^<$ and $G^r-G^a$ have sharp
$\delta$-function-like peaks at resonant level of the dot.

The Eq.~(\ref{eq:curr2}) is the main result of this paper. Note that the 
spectral function $\rho_{\sigma}(\omega)$ should be 
calculated {\em in the presence of the leads}, 
and it includes resonant tunneling, spin flips,
inelastic scattering, etc. 
For the normal metallic leads, $\Gamma_{\alpha}^S(\omega)
=\Gamma_{\alpha}$, and thus the Eq.
(\ref{eq:curr2}) 
is reduced to Meir and Wingreen's formula\cite{meir92}.
That is, the Eq.~(\ref{eq:curr2}) is a generallization of Meir and Wingreen's
formula to the case of the superconducting leads.

As an example of the quantum dot, let us consider an Anderson impurity.
The Hamiltonian of the dot is
\begin{equation}
 {\cal H}_D = \sum_{\sigma} \varepsilon_{\sigma} d_{\sigma}^{\dagger}
    d_{\sigma} + U n_{\uparrow}n_{\downarrow} , \label{eq:and_imp}
\end{equation}
where $n_{\sigma}=d_{\sigma}^{\dagger}d_{\sigma}$ is the number operator of 
the dot electron with spin $\sigma$. If one neglects higher order effects 
such as the Kondo-like correlations, the Green's function can 
be simply evaluated by the method of the 
equation of motion and truncation of 
high order Green's functions\cite{meir91,czycholl86,kang95}. 
To begin with, let us introduce $2\times2$ Nambu representation of the
(retarded) Green's function
\begin{equation}
 {\bf G}_{\sigma}(\omega) = \,\ll \bar{\gamma}_{\sigma}\; , \; 
    \bar{\gamma}_{\sigma}^{\dagger} \gg ,
\end{equation}
where 
$\bar{\gamma}_{\sigma} = \left( \begin{array}{c}
				 d_{\sigma} \\ d_{-\sigma}^{\dagger}
                               \end{array} \right)  .  $
The equation of motion for ${\bf G}_{\sigma}$ yields
\begin{eqnarray}
 &  \left( \begin{array}{cc}
	 \omega-\varepsilon_{\sigma} & 0 \\
	 0 & \omega+\varepsilon_{-\sigma}  \end{array} \right)
           {\bf G}_{\sigma}(\omega) &  \nonumber \\
 & = {\bf 1} + \sum_{k\in L,R} V_k \bar{\sigma}_z
     \ll \bar{\psi}_{k\sigma}\; , \; \bar{\gamma}_{\sigma}^{\dagger} \gg
     & \\
 &   + U\bar{\sigma}_z \ll \bar{\gamma}_{\sigma}^{(2)} \; , \;
     \bar{\gamma}_{\sigma}^{\dagger} \gg , & \nonumber
\end{eqnarray}
where $\bar{\sigma}_z = \left( \begin{array}{cc} 1 & 0 \\ 0 & -1 \end{array}
\right)$,
$\bar{\psi}_{k\sigma} = \left( \begin{array}{c} c_{k\sigma} \\ 
c_{-k-\sigma}^{\dagger} \end{array} \right) $, $\bar{\gamma}_{\sigma}^{(2)} = 
\left( \begin{array}{c} n_{-\sigma}d_{\sigma} \\
n_{\sigma}d_{-\sigma}^{\dagger} \end{array} \right)$
and ${\bf 1}$ is the unit matrix. This equation cannot be solved exactly due
to the last term of the R.H.S. In order to get an approximate solution of
${\bf G}_{\sigma}$, we consider the equation of motion for the Green's function
$\ll \bar{\gamma}_{\sigma}^{(2)} \; , \;
     \bar{\gamma}_{\sigma}^{\dagger} \gg$ and truncate higher order Green's
functions which appear on the equation of motion.
Adopting this procedure and taking $U\rightarrow\infty$ limit 
one can find that
\begin{equation}
 G_{\sigma}(\omega) = \left[ {\bf G}_{\sigma}(\omega) \right]_{11} \approx
 \frac{ 1-\langle n_{-\sigma}\rangle }{ \omega-\varepsilon_{\sigma}+
       i\left( \Gamma_L^S(\omega)+\Gamma_R^S(\omega) \right) } , 
\end{equation}
which leads to the resonant-tunneling-like expression of the current
\begin{eqnarray}
 I &=& \frac{4e}{h}\sum_{\sigma}\left( 1-\langle n_{-\sigma}\rangle \right) 
       \int d\omega  \left[ f_L(\omega)-f_R(\omega) \right]  \nonumber \\
   &\times&  \frac{ \Gamma_L^S(\omega)\Gamma_R^S(\omega) }{ 
	    \left( \omega-\varepsilon_{\sigma} \right)^2 +
	    \left( \Gamma_L^S(\omega)+\Gamma_R^S(\omega) \right)^2 }  
	     .  \label{eq:res}
\end{eqnarray}
The factor $\left( 1-\langle n_{-\sigma}\rangle \right)$ arises from 
strong Coulomb repulsion.
Similar expression to the Eq.~(\ref{eq:res}) has been obtained by Yeyati
{\em et al.}\cite{yeyati97}, where the dot Hamiltonian is simply simulated
by a single non-interacting level, 
instead of the Eq.~(\ref{eq:and_imp}).
%Here $\langle n_{\sigma}\rangle$ depends on the bias voltage. 
%In the $\Gamma_L,\Gamma_R\rightarrow 0$ limit, the 
%Eq.~(\ref{eq:res}) is reduced to
%an expression of the sequential tunneling current, 
%\begin{displaymath}
% I = \frac{2e}{\hbar}\sum_{\sigma}(1-\langle n_{-\sigma}\rangle )
% \tilde{\Gamma}^S(\varepsilon_{\sigma}) \left[ f_L(\varepsilon_{\sigma})
% - f_R(\varepsilon_{\sigma}) \right]  .
%\end{displaymath}
Resonance broadening of the spectral function
is included in the Eq.~(\ref{eq:res}). 
The effect of resonance broadening in the $I-V$ characteristic has been
discussed in the Ref.~\cite{yeyati97}.
It is notable that for a small resonance broadening $\Gamma_{\alpha}
\ll \Delta_{\alpha}$, the current via single level directly reflects
the BCS singularity in the density of states of the superconducting
lead via the function $\Gamma_{\alpha}(\omega)$ (see Fig.2(a)).

While the Eq.~(\ref{eq:res}) is valid for $\Delta_{\alpha} \gg 
\Gamma_{\alpha}$, Kondo-like correlation becomes important on
the other limit, $\Delta_{\alpha} \ll \Gamma_{\alpha}$. In order to describe 
the Kondo effect properly, more elaborate calculation, such as non-crossing
approximation (NCA)\cite{bickers87}, is required in calculating 
the spectral function of the dot electron. 
For normal metallic leads, NCA has been used
to investigate anomalies of the $I-V$ characteristic arising from the
Kondo resonances in finite bias voltages \cite{meir}. The formalism given
in this paper can address the problem of the non-equilibrium Kondo effect
in the presence of the superconducting leads. 
%Further study along this line is under progress.

Another interesting example is the boson-assisted transport. Consider a 
quantum dot in which the electrons interact with bosonic modes $\omega_q$
with the coupling strength $g_q$. The Hamiltonian of such a quantum dot
is given by
\begin{eqnarray}
 {\cal H}_D &=& \sum_{\sigma} \varepsilon_{\sigma} d_{\sigma}^{\dagger}
     d_{\sigma} + U n_{\uparrow}n_{\downarrow}
	\nonumber \\
  &+&
     \sum_q \hbar\omega_q b_q^{\dagger}b_q + \hat{n}\sum_q g_q (b_q+b_q^{\dagger})
     ,   \label{eq:el-bos} 
\end{eqnarray}
where $\hat{n}=\sum_{\sigma} n_{\sigma}$ and
$b_q^{\dagger}$ ($b_q$) creates (destroys) a boson.
For the full Hamiltonian ${\cal H}$ of the Eq.~(\ref{eq:hamil}) 
with the quantum dot
of the Eq.~(\ref{eq:el-bos}), a canonical transformation with 
$\phi = i\sum_q (g_q/\hbar\omega_q) ( b_q^{\dagger}-b_q )$ gives the new
effective Hamiltonian~\cite{konig96,mahan90} 
\begin{mathletters}
\begin{eqnarray}
 \bar{\cal H} &=& e^{-i\hat{n}\phi} {\cal H} e^{i\hat{n}\phi} \nonumber \\
    &=& {\cal H}_L + {\cal H}_R + \bar{\cal H}_D + \bar{\cal H}_T ,
\end{eqnarray}
where 
\begin{equation}
 \bar{\cal H}_D = \sum_{\sigma} \bar{\varepsilon}_{\sigma} n_{\sigma}
   + \bar{U} n_{\uparrow}n_{\downarrow} + \sum_q\hbar\omega_q b_q^{\dagger}b_q ,
\end{equation}
and 
\begin{equation}
 \bar{\cal H}_T = \sum_{k\sigma\in L,R} V_k \left( 
      c_{k\sigma}^{\dagger}d_{\sigma}e^{i\phi} +
      e^{-i\phi} d_{\sigma}^{\dagger}c_{k\sigma} \right).
\end{equation}
\end{mathletters}
The level position and the Coulomb energy in the dot are renormalized by
the electron-boson interaction as 
$\bar{\varepsilon}_{\sigma}=\varepsilon_{\sigma}
-\sum_q g_q^2/\hbar\omega_q$, $\bar{U}=U-2\sum_q g_q^2/\hbar\omega_q$.
The phase factors $e^{\pm i\phi}$ 
which appear in the tunneling term are related to the boson assisted tunneling.
With the transformed Hamiltonian $\bar{\cal H}$, 
the current formula (\ref{eq:curr2}) is modified as
\begin{equation}
 I = \frac{2e}{\hbar} \sum_{\sigma}\int d\omega\, \tilde{\Gamma}^S(\omega)
      \left\{ f_L(\omega)-f_R(\omega) \right\} \bar{\rho}_{\sigma}(\omega) ,
		     \label{eq:curr_el-bos}
\end{equation}
where $\bar{\rho}_{\sigma}$ is the spectral function of the composite particle
denoted by $\bar{d}_{\sigma}\equiv d_{\sigma}e^{i\phi}$ :
\begin{displaymath}
 \bar{\rho}_{\sigma}(\omega) = -\frac{1}{\pi}\, \mbox{Im} 
   \ll \bar{d}_{\sigma}\, , \,
   {\bar{d}_{\sigma}}^{\dagger} \gg .
\end{displaymath}
%For normal metallic leads, the Eq.~(\ref{eq:curr_el-bos}) has been 
%used to investigate boson-assisted transport\cite{konig96}.
%We would like to point out that the Eq.~(\ref{eq:curr_el-bos}) can be used
%The Eq.~(\ref{eq:curr_el-bos}) can be used
%to study interesting properties originating from emission and absorption
%of bosons in the presence of superconducting leads. 
%
%For example, one can describe multiple Kondo singularities 
%in the limit of $\Delta_{\alpha}=0$ \cite{konig96}, 
%and multiple BCS singularities in the
%$\Delta_{\alpha} \gg \Gamma_{\alpha}$ limit \cite{whan96}.
%

In general, exact calculation of $\bar{\rho}_{\sigma}(\omega)$ is not
possible. However, in the limit of weak coupling between the dot and
leads, that is for $\Gamma=\Gamma_L+\Gamma_R \ll k_B T$, the
leads can be neglected and the spectral function can be evaluated
exactly for that case. We consider the Einstein model, in which all
bosons have a same energy $\hbar\omega_0$ and are coupled to the
electrons in the quantum dot with strength $g_0$. In this model 
the spectral function is given by (see e.g. \cite{mahan90})
\begin{eqnarray}
 \bar{\rho}_{\sigma}(\omega) &=& e^{-\alpha(2N_0+1)} 
   \sum_{l=-\infty}^{\infty} I_l ( 2\alpha [N_0(N_0+1)]^{1/2} )
       \nonumber \\
    &\times& e^{\beta l\hbar\omega_0/2} 
    \left\{  \left( 1-\langle n_{-\sigma}\rangle \right) 
    \delta(\omega-\bar{\varepsilon}_{\sigma}-l\hbar\omega_0 ) \right.
	\label{eq:bat_rho} \\
    &+& \left. \langle n_{-\sigma} \rangle 
    \delta(\omega-\bar{\varepsilon}_{\sigma}-\bar{U}-l\hbar\omega_0 )
	  \right\} , \nonumber
\end{eqnarray}
where $N_0=1/(e^{\beta\hbar\omega_0}-1)$ and $\alpha=g_0^2/(\hbar\omega_0)^2$
are the boson occupation number and the dimensionless electron-boson 
coupling constant, respectively. The renormalized quantum dot level 
and the Coulomb repulsion
can be written as $\bar{\varepsilon}_{\sigma}=\varepsilon_{\sigma}
-\alpha\hbar\omega_0$, $\bar{U}=U-2\alpha\hbar\omega_0$. 
$I_l(x)$ is the modified
Bessel function. Applying the current formula (\ref{eq:curr_el-bos}) with
(\ref{eq:bat_rho}), one can get $I-V$ characteristics in the presence
of the electron-boson coupling.

Fig.2 shows the $I-V$ characteristics for different values of $\alpha$.
Here we consider the spin degenerate case ($\varepsilon_{\uparrow}=
\varepsilon_{\downarrow}=\varepsilon$). In the absence of the electron-boson
coupling (Fig.2(a)), the $I-V$ curve consists of two sharp peaks at
$eV=2(|\varepsilon|+\Delta)$ and at $eV=2(\varepsilon+U+\Delta)$. These
peaks reflect the singularity in the BCS density of states of the leads.
This feature, which makes better spectral resolution than the case
of the normal metallic leads, have been verified experimentally
\cite{ralph}.
Note that there is no thermal broadening in contrast to the case of the 
normal metallic leads. For normal metallic leads, there are two thermally
broadend steps at $eV=2|\varepsilon|, 2(\varepsilon+U)$. For $\alpha\ne0$,
the main peaks are located at $eV=2(|\bar{\varepsilon}|+\Delta), 
2(\bar{\varepsilon}+\bar{U}+\Delta)$. Further, we obtain additional 
side peaks at $eV=2(|\bar{\varepsilon}+l\hbar\omega_0|+\Delta)$ and
at $eV=2(\bar{\varepsilon}+\bar{U}+l\hbar\omega_0+\Delta)$ with
positive integer $l$ related to boson emission. Boson absorption processes 
are negligible because $k_BT\ll\hbar\omega_0$ is taken here.
For weak electron-boson coupling, only single boson emission process
contributes to the $I-V$ curve (Fig2.(b)). As the coupling becomes stronger,
multiple side peaks appear (Fig2.(c)). For normal metallic leads, side
steps arise from the boson assisted tunneling, but those are not clearly
shown because of thermal broadening.

In conclusion, we have derived a formula for the current through an interacting
quantum dot coupled to two suerconducting leads, when the Andreev reflecction
processes are negligible because of the strong Coulomb repulsion in the dot. 
This formula provides a new framework to study 
interaction effects in the transport through mesoscopic systems coupled to
superconductors. For example,
quasiparticle resonant tunneling has been described by an equation of
motion method in the limit of large Coulomb repulsion in the dot. 
%Further investigations including higher order effects are under study.
%Further, we have discussed several problems which can be treated by
%this new formulation, such as the Kondo-resonant transport and 
%the boson assisted tunneling.
Further, current-voltage charactersistics of boson assisted tunneling
have been studied. Splitting of BCS singularities, which arises from the
boson assisted tunneling, has been shown in the $I-V$ characteristics.

The author thanks B.I. Min for useful discussions.
This work has been supported by the Korea Science and
Engineering Foundation.

%%%%%%%%%%% References %%%%%%%%%%%%%%%%%%%%%%%%%%%%%%%%%%%%%%%%%%%%%%%

%
\end{multicols}
\begin{figure}
 \caption{ Schematic diagram of a quantum dot connected to two superconducting
 leads. The superconducting leads are characterized by the chemical potentials
 $\mu_L,\mu_R$ and the energy gaps $\Delta_L,\Delta_R$.
 An eletron current $I$ flows from left to right if $\mu_L-\mu_R=eV > 0$.
	   }
\end{figure}
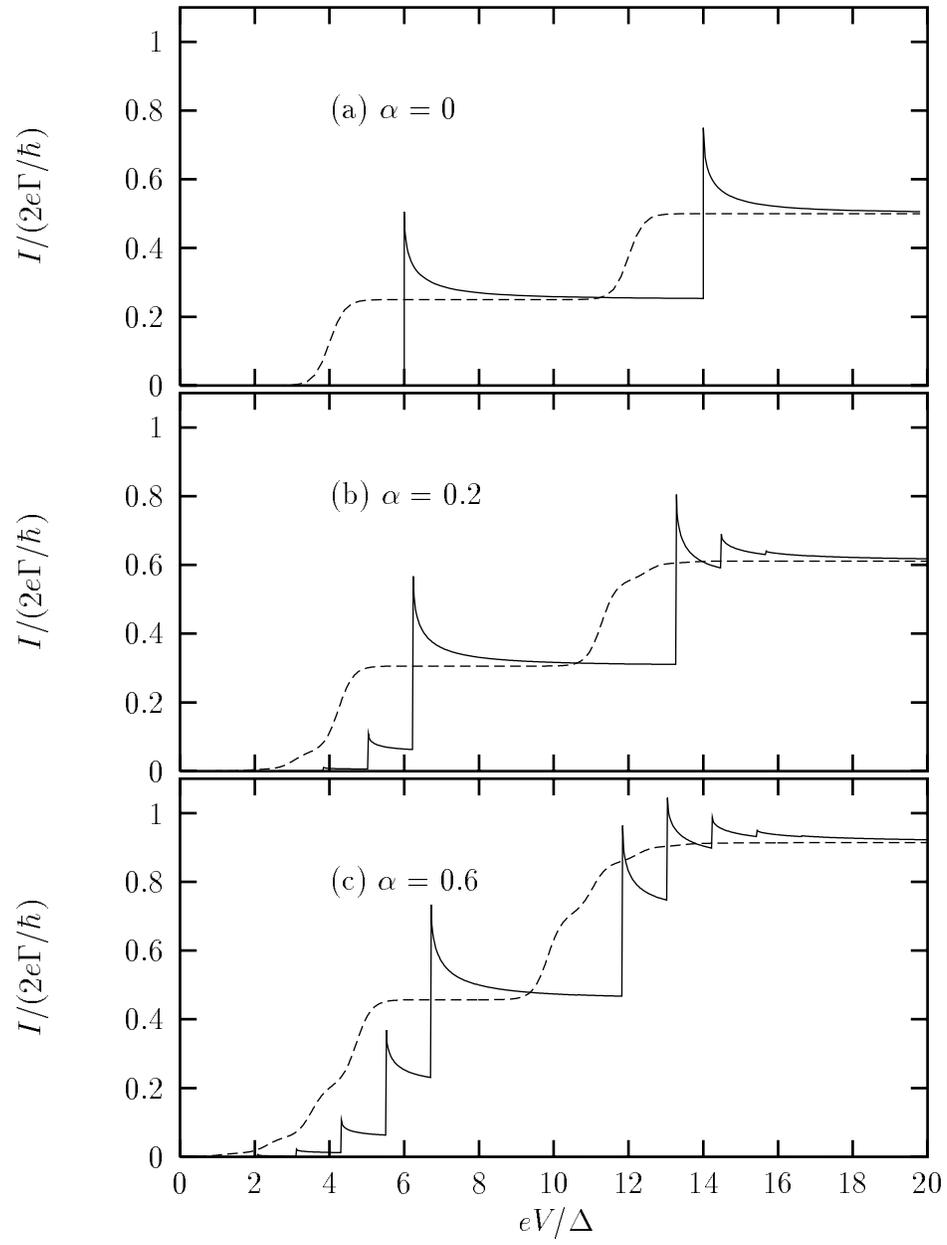
\begin{figure}
 \caption{ Current-voltage characteristics (a) without, and (b)(c) with boson
 assisted tunneling. Here $\Delta_L=\Delta_R=\Delta$, $\Gamma_L=\Gamma_R
 =\Gamma/2$ and $\mu_L=eV/2$, $\mu_R=-eV/2$ are considered. Other parameters
 used here are $k_BT=0.1$, $\varepsilon=\varepsilon_{\uparrow}
 =\varepsilon_{\downarrow}=-2$, $U=8$, $\hbar\omega_0=0.6$ in the unit
 of $\Delta$.  Full lines represent the results for the superconducting 
 leads, while dashed lines being corresponding results for the normal 
 metallic leads.
  }
\end{figure}
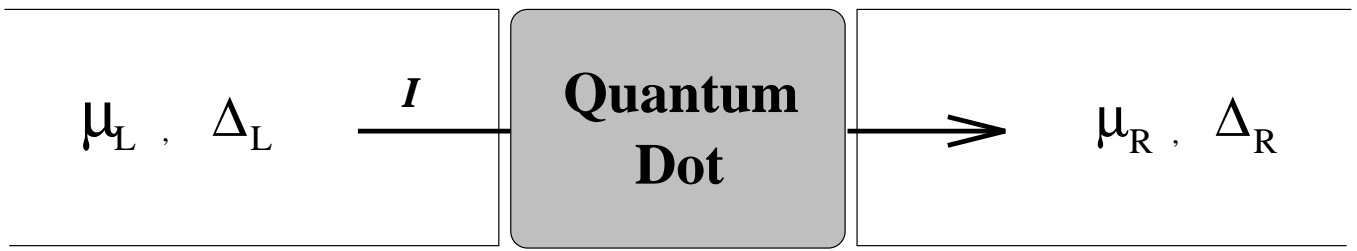
\end{document}